\documentstyle[twocolumn,aps,prl,epsf]{revtex}
\begin{document}
\title{Remnant Fermi Surfaces in Photoemission}

\author{C. Kusko$^*$ and R.S. Markiewicz} 

\address{Physics Department and Barnett Institute, 
Northeastern U.,
Boston MA 02115}
\maketitle
\begin{abstract}
Recent experiments have introduced a new concept for analyzing the 
photoemission spectra of correlated electrons -- the remnant Fermi surface 
(rFs), which can be measured even in systems which lack a conventional Fermi 
surface.  
Here, we analyze the rFs in a number of interacting electron models, and find
that the results fall into two classes.  For systems with pairing instabilities,
the rFs is an accurate replica of the true Fermi surface.  In the presence of
nesting instabilities, the rFs is a map of the resulting superlattice Brillouin
zone.  The results suggest that the gap in Ca$_2$CuO$_2$Cl$_2$ is of nesting
origin.
\end{abstract}


\narrowtext

Recently, a new experimental tool has been introduced\cite{Shen} to 
parametrize photoemission (PE) data in strongly correlated metals: the `remnant 
Fermi surface' (rFs).  This is the locus of points in $\vec k$-space where the 
PE intensity associated with a particular quasiparticle peak falls to half of 
its peak value.  For an ordinary metal, these points would correspond to the 
true Fermi surface, but in strongly correlated metals the points do not
necessarily fall at the same energy -- the rFs may display a considerable
dispersion.  

Ronning, et al.\cite{Shen} measured the rFs of
Ca$_2$CuO$_2$Cl$_2$ (CCOC), a half filled Mott insulator, and compared it with
the rFs of optimally doped Bi$_2$Sr$_2$CaCu$_2$O$_8$ (BSCCO).  When underdoped, 
the dispersion of BSCCO evolves toward that of CCOC\cite{Laugh1}.  
Qualitatively, the rFs of CCOC seems consistent with Luttinger's theorem, even 
though it displays a considerable dispersion. Despite this, the rFs's of the two
materials are strikingly different, and cannot evolve from each other via rigid
band filling (they cross).  A proper understanding of the rFs could lead to an 
improved model for the pseudogap in these materials.
 
In this Letter we analyze the rFs expected for a variety of interacting electron
systems, and show that they do not necessarily provide information about the 
`true' Fermi surface.  The results fall into two classes, depending on whether 
the interaction can be characterized as `nesting' or `pairing'.  Only in the 
latter case is the rFs a reliable map of the Fermi surface.  In the former case,
it maps out the superlattice Brillouin zone generated by the nesting 
instability.
\par 
A number of different mechanisms have been proposed for the origin of pseudogaps
in the cuprates.  These include magnetic (spin density waves)\cite{KaSch}, flux 
phase (RVB)\cite{Laugh,WeL,Pstr,CZ}, charge ordering (CDW)\cite{Pstr,RAK}, and 
superconducting fluctuations\cite{Rand}.  These instabilities 
fall into two classes nesting (the first three: associated with particle-hole
propagators, and instabilities in the charge or spin susceptibilities at a
nesting vector, here $\vec Q=(\pi ,\pi )$) and pairing (the last, associated
with particle-particle propagators and the uniform susceptibility at $\vec 
q=0$).  In the cuprates, {\it all} of these instabilities may be analyzed within
an SO(6) group -- the instability group of the Van Hove singularity\cite{SO6}.
We find that the rFs has two strikingly different origins in these two classes, 
but there is relatively little variation within a given class.  In CCOC the rFs
seems to indicate the locus of the reduced Brillouin zone, suggesting a nesting 
instability.
\par
An important sum rule for ARPES that relates the integrated intensity to the 
momentum distribution has been introduced by Randeria et al.\cite{Rand1} and is 
given by
\begin{equation}
n(\vec k)=\int_{-\infty}^{\infty}{d\omega}f(\omega)A(\vec k,\omega),
\label{eq:1}
\end{equation} 
where $A(\vec k,\omega)$ is the one particle spectral function of the model,
$n(k)=<c_{\vec k}^{\dagger}c_{\vec k}>$
is the momentum distribution and $f(\omega)$ is the Fermi function.
Although $n(\vec k)$ is a ground state property, they proved that in
the limit of the sudden approximation the frequency integrated
spectral function gives the momentum distribution. They employed this 
sum rule to determine the momentum distribution in
BSSCO and YBCO. Ronning et al.\cite{Shen} extended this methodology to strongly
correlated electron compounds. By defining $k_F$ as the point of
steepest descent, they showed that even when strong Coulomb correlations 
destroy the Fermi-liquid character of the system, $n(\vec k)$ still drops 
sharply, allowing the determination of a rFs.
\par
\begin{figure}
\leavevmode
   \epsfxsize=0.45\textwidth\epsfbox{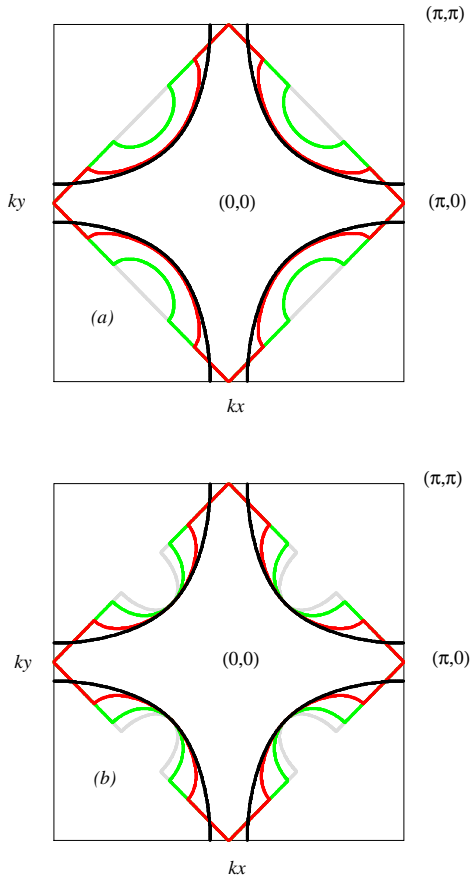}
\vskip0.5cm 
\caption{Remnant Fermi surfaces for different nesting
instabilities at $T$=0K. (a) evolution of the rFs toward a perfect square with
increasing CDW gap; (from darkest to lightest) $O_{CDW}$=0, 100, 300, 500
meV. (b) the same evolution for a flux phase instability.  Superconducting
instabilities are not shown, since in this figure the rFs would coincide with 
the true Fermi surface.}
\label{fig:1}
\end{figure}
\begin{figure}
\leavevmode
   \epsfxsize=0.4\textwidth\epsfbox{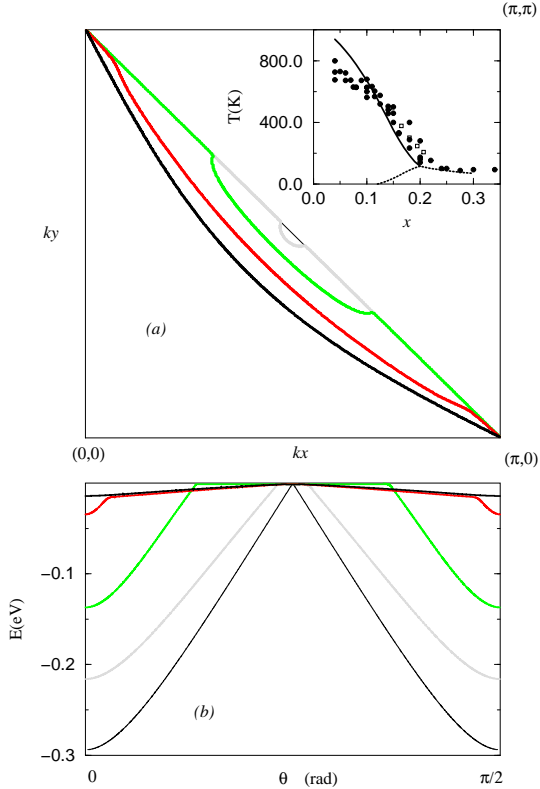}
\vskip0.5cm 
\caption{(a) Evolution of the rFs from half filling to optimal doping
at $T=0K$ with self-consistent gap parameters
for a model with d-wave superconductivity and flux phase. 
Curves from ligthest to darkest:
$x$=0 (thin line), 0.04, 0.1, 0.19 and 0.26 (black  line).  
Inset: Pseudogap phase diagram for LSCO (filled circles) and YBCO
(open squares) determined from transport\protect\cite{Batl}. 
Solid line - $O^{JC}$ transition
$T^*$; dotted line - superconductivity transition $T_c$. 
(b) quasiparticle dispersion along the rFs plotted in (a), 
$\tan\theta=k_x/k_y$}
\label{fig:2}
\end{figure}

In the following, we calculate the spectral function $A(\vec k,\omega)$
and the momentum distribution for mean field models with a variety of 
instabilities.  Figure~\ref{fig:1} illustrates the rFs's associated with 
several nesting instabilities.  The energy dispersion has the standard one-band 
form
\begin{equation}
e_{\vec k}=-2t_0(\cos{k_xa}+\cos{k_ya})-4t_1\cos{k_xa}\cos{k_ya},
\label{eq:1a}
\end{equation} 
with $t_0=0.25eV$, $t_1=-0.45 t_0$.  The calculations follow those in
Ref.~\cite{MKK} for CDW and s-wave superconductivity.  The rFs's for all the
nesting instabilities are essentially identical: over part of the surface, there
is no gap, and the rFs is dispersionless, coinciding with the true Fermi 
surface at $e_{\vec k}=E_F$; here $n(\vec k)=1/2$ is mainly due to
the Fermi function in Eq.~\ref{eq:1}.  Over the rest of the zone, the rFs lies
along the zone diagonals which determine the $\sqrt{2}\times\sqrt{2}$ nesting
superlattice.  On this part of the rFs there is considerable dispersion, and
$n(\vec k)=1/2$ due to the coherence factor, discussed below.  By contrast, the
rFs for a pairing instability is always located below $E_F$ at the 
superconducting gap, dispersionless for s-wave, dispersive for d-wave, but in
both cases faithfully following the contours of the true Fermi surface.
If the pseudogap is due to a nesting instability competing with 
superconductivity, then there should be a characteristic evolution of the rFs
with doping, from nesting-like at half filling to pairing-like in the overdoped
regime.  The phase diagram has been worked out for such a competition, both
for CDW-to-s-wave superconductivity\cite{MKK} and for flux phase to 
d-wave\cite{CZ}.  In both cases, we find the evolution of the rFs's is nearly
identical.  Figure~\ref{fig:2} illustrates this evolution for the latter case.
Note that since the phase at half filling is fully gapped (a Mott insulator),
the rFs is perfectly square.  The two limiting cases, insulator and optimally
doped, bear a marked resemblance to the experimental observations\cite{Shen}.

In the calculations of Fig.~\ref{fig:2}, we took the competing phases to be
d-wave superconductivity, with gap 
$\Delta_{\vec k}^{d}=\Delta^{d}\gamma_{\vec k}$, 
with $\gamma_{\vec k}=\cos k_xa-\cos k_ya$ and an orbital
antiferromagnet
\cite{Sch2,SO6}, a nesting 
instability with gap $O_{\vec k}^{JC}=O^{JC}\gamma_{\vec k}$, which is 
an RVB state having a d-wave symmetry corresponding to a particle-hole 
excitation, essentially equivalent to the flux phase instability introduced by 
Affleck and Marston\cite{Affl}. We consider a one-band model, Eq.~\ref{eq:1a},
with correlation effects simulated by a doping dependent $t_0=xt_0^*$, 
$t_0^*=2.3eV$, and 
\begin{equation}
t_1/t_0=-0.52\tanh(2.4x), 
\label{eq:1b}
\end{equation} 
to pin the Van Hove singularity (VHS) close to the Fermi level over an
extended range of doping\cite{Pstr,MKK}. 
We start with the following mean-field
hamiltonian,
\begin{eqnarray}
H=\sum_{\vec k,\sigma}\epsilon_{\vec k}c_{\vec k
\sigma}^{\dagger}c_{\vec k \sigma}
+O_{\vec k}^{JC}(c_{\vec k\sigma}^{\dagger}c_{\vec k+\vec Q
\sigma}+h.c.)+\nonumber \\
\sum_{\vec k}\Delta_{\vec k}^{d}(c_{\vec k \uparrow}^{\dagger}c_{-\vec k
\downarrow}^{\dagger}+h.c.).
\end{eqnarray}   
with the quasiparticle dispersion given by,
\begin{equation}
E_{\pm, \vec k}^2={1 \over 2}(\epsilon_{\vec k}^2+\epsilon_{\vec k+\vec
Q}+2\Delta_{\vec k}^2+{O_{\vec k}^{JC}}^2\pm\hat E_{\vec k}^2))
\end{equation}
Performing a generalized Bogoliubov-Valatin transformation we find
the following gap equations:
\begin{eqnarray}
\Delta^d=U_\Delta\sum_{\vec k}{{\gamma_{\vec k}}\over 2}
({\cos^2\phi}\sin2\phi_+
\tanh{{\beta E_{+, \vec k}}\over2}+
\nonumber \\
{\sin^2\phi}\sin2\phi_-
\tanh{{\beta E_{-, \vec k}}\over2})
\end{eqnarray}
and
\begin{eqnarray}
O^{JC}=U_{O^{JC}}\sum_{\vec k}{{\gamma_{\vec k}}\over 4}
\sin2\phi(\cos2\phi_{+}\tanh{{\beta E_{+,\vec k}}\over2}+
\nonumber \\
\cos2\phi_{-}\tanh{{\beta E_{-,\vec k}}\over2}) 
\end{eqnarray}
with $\hat E_{\vec k}=\sqrt{(\epsilon_{\vec k}-\epsilon_{\vec k+\vec
Q})^2+4{O_{\vec k}^{JC}}^2}$ and the angles defined by 
$\tan2\phi_\pm=2\Delta_{\vec k}
/(\epsilon_{\vec k}+\epsilon_{\vec k+\vec Q}\pm\hat E_{\vec k})$, 
\begin{equation}
\cos2\phi=\frac{\epsilon_{\vec k}-\epsilon_{\vec k+\vec Q}}{\hat E_{\vec k}},
\end{equation}
and $\epsilon_{\vec k}=e_{\vec k}-E_F$. 
Solving the gap equations self-consistently with the coupling
constants $U_{O^{JC}}$=176 meV and $U_\Delta$=88 meV, we find that
the 
experimental
phase diagram can be fit semi-quantitatively by this model (insert in 
Fig.~\ref{fig:2}) \cite{Batl}.  This phase diagram differs from that found in 
Ref.~\cite{MKK}, but not because different instabilities were assumed.  The main
difference arises from the gap cutoffs: in Ref.~\cite{MKK}, a phononic 
interaction was assumed, with cutoff $\hbar\omega =45meV$; here, the interaction
is taken as electronic, with no cutoff.  This makes a significant difference in
the phase boundary near half filling.

In the above calculations, the shape of the Fermi surface evolved with doping,
to mimic the effect of Van Hove pinning\cite{RM3,Pstr}.  However, the evolution
of the rFs for a nesting gap is generically toward a square Fermi
surface as the gap is increased, as illustrated in Fig.1.  The rFs has two
parts, a true (hole pocket) Fermi surface on the ungapped part of the rFs, and
a segment of square on the gapped part.  With increasing gap magnitude, the
former feature shrinks and the latter grows, until the full Fermi surface is
gapped and the rFs is square. Note that due to the d-wave symmetry the
rFs for a flux phase instability is pinned at the true Fermi surface
along the diagonal $(0,0) - (\pi,\pi)$, Fig.1b.  
The shrinking of the true Fermi surface is
reminiscent of the evolution in BSCCO reported by Norman, et al.\cite{NDR}.
It should be noted that the rFs is not equivalent to the minimum gap locus
introduced by Ding, et al.\cite{DgNY}.

The origin of the rFs can be understood from these calculations.  In the
competing flux phase-d-wave model, $n(\vec k)$ can be written as
\begin{eqnarray}
n(k)={1\over 2}(1-\cos ^2\phi\cos2\phi_+\tanh{{\beta E_{+,\vec k}\over 2}}-
\nonumber \\
\sin^2\phi\cos2\phi_-\tanh{{\beta E_{-, \vec k}\over 2}}).
\end{eqnarray}
For a pure d-wave superconductivity model this becomes
\begin{equation}
n(k)={1\over 2}(1-{\epsilon_{\vec k}\over E_{\vec k}}
\tanh{{\beta E_{\vec k}\over 2}}),
\end{equation}
with $E_{\vec k}=\sqrt{\epsilon_{\vec k}^2+\Delta_{\vec k}^2}$, 
showing that the rFs coincides with the true Fermi surface:
$n(\vec k)$=1/2 when $\epsilon_{\vec k}=0$.
For a pure nesting model $n(\vec k)$ is given by
\begin{eqnarray}
n(\vec k)={1\over 2}(1-\cos ^2\phi\tanh{{\beta E_{+,\vec k}^{nest}\over 2}}-
\nonumber \\
\sin^2\phi\tanh{{\beta E_{-, \vec k}^{nest}\over 2}}),
\end{eqnarray}
with $E_{\pm, \vec k}^{nest}=(\epsilon_{\vec k}+\epsilon_{\vec k+\vec Q}\pm\hat
E_{\vec k})/2$.
As $T \rightarrow 0$, the two $\tanh$'s go to 1 or -1, 
so $n(\vec k)=1/2$ when $\cos^2\phi-\sin^2\phi=0$, or, from Eq.[8],
$\epsilon_{\vec k}=\epsilon_{\vec k+\vec Q}$. For the present model,
this is the superlattice Brillouin zone boundary.  
\par
In the underdoped regime, as temperature is lowered the cuprates pass 
first into the pseudogap phase, at temperature $T^*$, 
then into a superconducting phase at 
$T_c$.  In the present scenario, $T^*$ would signal a transition to a nested
phase with a gap (or pseudogap if realistic fluctuations are 
included\cite{KaSch,RM5}), leaving hole pockets behind.  Below $T_c$, an 
additional, pairing gap opens at the hole pockets.  However, a careful look at
the rFs shows a more complicated evolution, Fig.3a: the shape of the
hole pockets changes, with an accompanying transfer of spectral weight from the
nesting to the pairing parts of the rFs.
Note that in Fig.3a the rFs has the same locus in k-space
as the true hole pocket Fermi surface above Tc, but from
Fig.3b there is a dramatic shift in dispersion of this rFs as
the superconductivity gap opens.

In comparing these results to experiment, the rFs of CCOC clearly displays the
square shape expected for a predominantly nesting interaction.  This is 
consistent with all of the pseudogap models noted above, except for preformed
pairs.  In fact, preformed pairs would still be a possibility, if strong
correlation effects renormalized the (true) Fermi surface to square at half
filling.  Such renormalization has been proposed previously\cite{RM3,Surv1},
and is incorporated in Eq.~\ref{eq:1b}.  However, in these theories, the
renormalization leads to greatly enhanced nesting, and is less favorable for
pairing.  The most likely conclusion is that the pseudogap in the underdoped
cuprates represents some nesting instability, which is fundamentally competing
with superconductivity.  Clearly, since the cuprates are 
quasi-two-dimensional, 
there should be prominent superconducting fluctuations above $T_c$, but they do
not represent the dominant part of the pseudogap.
\begin{figure}
\leavevmode
   \epsfxsize=0.4\textwidth\epsfbox{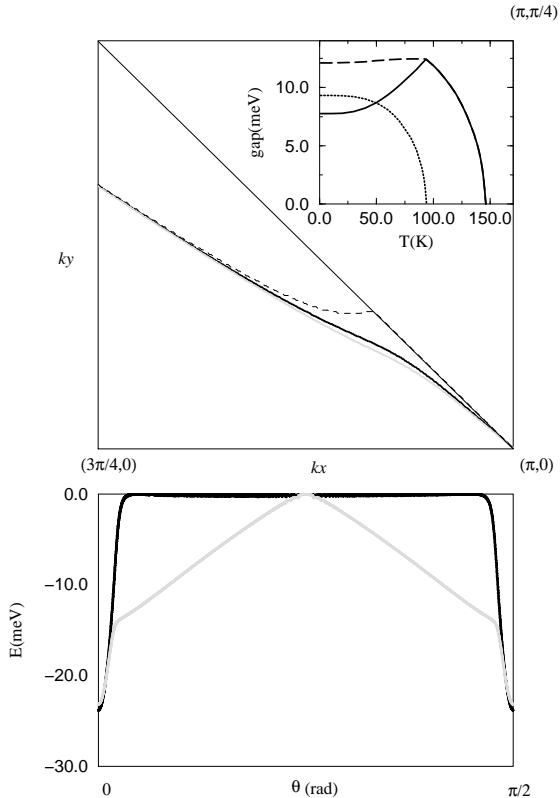}
\vskip0.5cm 
\caption{ (a) Evolution of the rFs with temperature for a fixed
doping $x$=0.19: black line - $T$=0K,
grey line - $T=T_c$=94K ($\Delta^d$=0 meV), dashed line shows what
the flux phase rFs at T=0K would be if  $\Delta^d$=0 meV. Inset:
Temperature dependence of the superconducting and flux phase gaps; dotted
line - $\Delta^d$, solid line - $O^{JC}$, dashed line
- $\protect\sqrt{\Delta^{d2}+O^{JC2}}$
(b) quasiparticle dispersion along the rFs plotted in (a).}
\label{fig:3}
\end{figure}
\par
The present results suggest a number of experimental tests.  The rFs should 
be mapped out in the cuprates as a function of doping.  In particular, the 
results of Norman, et al.\cite{NDR} should be extended to the full rFs.
Observation of a shift in spectral weight with temperature, Fig.~\ref{fig:3}
would provide strong evidence that the pseudogap is a nesting phenomenon, and
not due to preformed pairs.
Moreover, the rFs can be studied in other systems, to confirm the predicted
properties.  A start has already been made in CDW systems\cite{CDW}.  Mott
insulators would be particularly of interest.  It is believed that the 
insulating phase can form in the absence of magnetic order, hence without
nesting.  In the cuprates, there is a clear N\'eel transition, so a square rFs
is not unexpected, but a study of rFs's in non-magnetic Mott insulators could 
prove most informative.
\newline
\newline
Publication 760 of the Barnett Institute.
\newline
*: On leave of absence from Institute of Atomic Physics, Bucharest, Romania.

\end{document}